\documentclass[sigconf]{acmart}

\copyrightyear{2026}
\acmYear{2026}
\setcopyright{cc}
\setcctype{by}
\acmConference[WWW '26] {Proceedings of the ACM Web Conference 2026}{April 13--17, 2026}{Dubai, United Arab Emirates.}
\acmBooktitle{Proceedings of the ACM Web Conference 2026 (WWW '26), April 13--17, 2026, Dubai, United Arab Emirates}
\acmISBN{979-8-4007-2307-0/2026/04}
\acmDOI{10.1145/3774904.3792968}
\usepackage{comment}
\usepackage{multirow}
\usepackage{amsmath,pifont}
\usepackage{booktabs}
\usepackage{rotating}
\usepackage{array}
\usepackage[para]{footmisc}
\usepackage{xcolor}
\usepackage{colortbl}
\usepackage{longtable}
\usepackage{circledsteps}
\usepackage{romannum}
\usepackage{tcolorbox}
\usepackage{enumitem} 
\usepackage{balance}
\usepackage{bm}
\usepackage{listings}

\definecolor{Ruling}{HTML}{18AFCC}
\definecolor{Request}{HTML}{5F9099}
\definecolor{Construct}{HTML}{CC1865}

\definecolor{Yes}{HTML}{64FFA0}
\definecolor{Part}{HTML}{FF9963}
\definecolor{No}{HTML}{FF6B63}

\newcommand{\cmark}{\cellcolor{Yes}\textcolor{black}{\ding{51}}}
\newcommand{\pmark}{\cellcolor{Part}\textcolor{gray}{$\bm{\approx}$}}
\newcommand{\xmark}{\cellcolor{No}\textcolor{white}{\ding{55}}}
\newcommand{\namark}{\cellcolor{white}\textcolor{black}{n/a}}

\lstdefinelanguage{Turtle}{
  keywords={a, rdf, rdfs, owl, foaf, xsd, base, prefix, mg, log, list},
  sensitive=true,
  morecomment=[l]{\#}, 
  morestring=[b]",
  morestring=[b]',
  alsoletter={\@}, 
  commentstyle=\color{gray},
  keywordstyle=\color{blue},
  stringstyle=\color{red},
}

\begin{document}

\title{"They've Stolen My GPL-Licensed Model!": Toward Standardized and Transparent Model Licensing}

\author{Moming Duan}
\orcid{0000-0001-5402-6634}
\affiliation{%
  \institution{East China Normal University}
  \city{Shanghai}
  \country{China}}
\email{mmduan@dase.ecnu.edu.cn}

\author{Rui Zhao}
\orcid{0000-0003-2993-2023}
\affiliation{%
  \institution{University of Oxford}
  \city{Oxford}
  \country{UK}}
\email{rui.zhao@cs.ox.ac.uk}

\author{Linshan Jiang}
\orcid{0000-0001-8501-9488}
\affiliation{%
  \institution{National University of Singapore}
  \city{Singapore}
  \country{Singapore}}
\email{linshan@nus.edu.sg}

\author{Nigel Shadbolt}
\orcid{0000-0002-5085-9724}
\affiliation{%
  \institution{University of Oxford}
  \city{Oxford}
  \country{UK}}
\email{nigel.shadbolt@cs.ox.ac.uk}

\author{Bingsheng He}
\orcid{0000-0001-8618-4581}
\affiliation{%
  \institution{National University of Singapore}
  \city{Singapore}
  \country{Singapore}}
\email{hebs@comp.nus.edu.sg}


\begin{abstract}
As model parameter sizes scale into the billions and training consumes zettaFLOPs of computation, the reuse of Machine Learning (ML) assets and collaborative development have become increasingly prevalent in the ML community.
These ML assets, including models, datasets, and software, may originate from various sources and be published under different licenses, which govern the use and distribution of licensed works and their derivatives.
However, commonly chosen licenses, such as GPL and Apache, are software-specific and are not clearly defined or bounded in the context of model publishing.
Meanwhile, the reused assets may also be under free-content licenses and model licenses, which pose a potential risk of license noncompliance and rights infringement within the model production workflow.
In this paper, we address these challenges along two lines:
1) For ML workflow compliance, we propose ModelGo (MG) Analyzer, a tool that incorporates a vocabulary for ML workflow management and encoded license rules, enabling ontological reasoning to analyze rights granting and compliance issues.
2) For standardized model publishing, we introduce ModelGo Licenses, a set of modell-specific licenses that provide flexible options to meet the diverse needs of the ML community.
MG Analyzer is built on Turtle language and Notation3 reasoning engine, envisioned as a first step toward Linked Open Data for ML workflow management.
We have also encoded our proposed model licenses into rules and demonstrated the effects of GPL and other commonly used licenses in model publishing, along with the flexibility advantages of our licenses, through comparisons and experiments.

\end{abstract}




\maketitle
\newcommand\webconfavailabilityurl{https://doi.org/10.5281/zenodo.18277024}
\ifdefempty{\webconfavailabilityurl}{}{
\begingroup\small\noindent\raggedright\textbf{Resource Availability:}\\
The source code of this paper has been made publicly available at \url{\webconfavailabilityurl}.
\endgroup
}

\section{Introduction}
\label{sec:intro}
In recent years, the compelling generalization capabilities provided by billion-parameter models~\cite{wei2023inverse}, along with the high computational and data costs associated with their training~\cite{maslej2024aiindex}, have motivated ML developers to collaborate incrementally rather than train models from scratch.
A common approach is to download a Pre-Trained Model (PTM)~\cite{jiang2023empirical} and fine-tune it for downstream task~\cite{hu2022lora}.
However, these practices can pose legal risks if the use or redistribution of reused components violates their licenses, akin to GPL violations in Open Source Software (OSS)~\cite{mathur2012empirical}.
Failure to comply with license terms can result in termination and legal action\footnote{See CoKinetic Systems, Corp. \textit{v}. Panasonic Avionics Corporation, 1:17-cv-01527.}.
Another concern is the license chosen for republishing the work.
Some developers adhere to traditional software publishing practices and select OSS licenses for their models~\cite{devlin2019bert,ni2022expanding, duan2025position}, which often lack clear definitions and conditions regarding ML activities and do not effectively prevent undesirable use.
For example, a licensee can close-source your published models, even if they are licensed under GPL, without violating any terms.

There are three possible ways to mitigate above risks.
First, developers could avoid using any third-party materials. However, this is extremely difficult for individual developers, as training PTMs is expensive and requires vast amounts of data. For instance, the training dataset for GPT-2~\cite{radford2019language} was collected from 45 million web pages, governed by various licenses and terms of use.
Second, a new publishing standard for ML projects could be developed, which might include drafting specific licenses for models and datasets~\cite{benjamin2019towards,contractor2022behavioral}, along with a compatibility table to guide their reuse policies. 
However, this approach does little to address existing conflicts in ML projects that already rely on assets released under traditional licenses. 
Furthermore, it is impractical to expect all publishers to relicense their previous works.
Third, we can scan the reused assets in ML projects and analyze existing license noncompliance to eliminate them. 
This is a common solution applied to OSS projects~\cite{jaeger2017fossology} but it cannot be directly extended to ML projects, as they can involve complex coupling mechanisms and different licensing frameworks that are interwoven within an ML development workflow.

Take MixLoRA~\cite{li2024mixlora} as an example: it is licensed under Apache-2.0 (an OSS license) and is fine-tuned on Llama 2 model~\cite{touvron2023llama} (governed by Llama 2 Community License~\cite{meta2024llama2}, a model license) using the Cleaned Alpaca Dataset~\cite{taori2023alpaca}, which is licensed under CC BY-NC-4.0 (a free-content license from Creative Commons, a.k.a CC~\cite{creative2024list}).
Previous OSS license analysis tools~\cite{mathur2012empirical,ombredanne2020free} that only consider package reference dependencies and focus on software licenses will fall short in such ML scenarios.
Therefore, to provide license analysis for ML projects, the key is to develop a specific ontology that describes the ML workflow and provides a corresponding interpretative solution for licenses, covering all licensing frameworks and disambiguating their mapping rules related to ML activities.
Moreover, the lack of consensus in standard model publishing practices and the inflexibility of existing publicly available model licenses have led many developers to publish their models under OSS licenses or even free-content licenses~\cite{groeneveld2024olmo,cohereforai2024c4ai}, further complicating the design of license analysis methods.

In this paper, we propose a two-pronged approach to address these challenges. 
First, to resolve existing noncompliance in ML projects, we introduce \textit{MG Analyzer}, a tool that constructs ML workflows as Resource Description Framework (RDF)~\cite{pan2009resource} graphs and assesses potential license compliance issues, improper license selection, granting of rights, restrictions, and obligations related to ML asset reuse. 
Second, we propose \textit{MG Licenses}, a new set of model-specific licenses with CC-style licensing options, to promote standardized model publishing.
To present potential risks of using OSS, model, and free-content licenses in model publishing scenarios, we evaluate them with the \emph{MG Analyzer} on a typical workflow.
We also demonstrate the flexibility of \textit{MG Licenses} in encompassing nearly all licensing conditions provided by other model licenses through comparisons.
The main contributions of our paper are:

\begin{itemize}
    \item We identified the challenges of license compliance and model licensing in ML workflow management.
    
    \item We developed MG Analyzer to automate license analysis in ML projects. It includes a vocabulary for describing ML workflows with dependencies and license rules, and provides an interface to convert user-input workflow descriptions into RDF graphs. These graphs enable the tool to construct dependencies, reason, and detect license conflicts.
    \item We drafted a set of model licenses called MG Licenses to promote more standardized model publishing. These licenses are well-defined and cover a complete spectrum of model publishing scenarios. Furthermore, we have integrated support for MG Licenses within MG Analyzer.
\end{itemize}



\section{Background and Related Work}
\label{sec:related}

\subsection{ML Workflow Management}
Accurate description and management of the provenance and dependencies of ML assets are critical for legal compliance analysis in ML workflows~\cite{wang2025ml}. 
Currently, there are two active directions toward standardized ML workflow management. 
The first is the development of AI Bills of Materials (BOM) formal description frameworks designed to enumerate ML assets, parameters, and procedures required to develop AI models.
Among these efforts, Software Package Data Exchange 3.0 (SPDX 3.0) stands out as an open standard for expressing AI BOMs~\cite{karen2024implementing}. 
SPDX 3.0 provides a graph-based, extensible data model that enables detailed provenance tracking and license metadata across complex software and AI development pipelines.
CycloneDX\footnote{\href{https://cyclonedx.org/}{CycloneDX Bill of Materials Standard}} is another emerging standard focused on cybersecurity and supply chain risk, supporting metadata for vulnerability disclosure, threat modeling, and policy enforcement.
While these standards offer a strong foundation for traceability and transparency, they remain general-purpose and lack native support for domain-specific semantics required in ML legal compliance analysis. 
In particular, they do not provide built-in vocabulary for ML-specific development actions such as distillation, fine-tuning, or embedding, nor do they capture legal concepts such as granted rights, reserved rights, and license conditions in a machine-interpretable form.

The second is the development of MLOps toolkits, such as ModelDB~\cite{vartak2018modeldb}, MLflow~\cite{zaharia2018accelerating}, Texera~\cite{wang2024texera, huang2024demonstration}, Neptune\footnote{\href{https://neptune.ai/}{neptune.ai}}, and SAS Model Manager\footnote{\href{https://www.sas.com/en_us/software/model-manager.html}{SAS Model Manager}}. 
These platforms facilitate the end-to-end management of ML workflows by supporting functionalities such as experiment tracking, model versioning, performance monitoring, and deployment orchestration. 
While they enhance reproducibility, governance, and automation in ML lifecycle management, they typically lack native support for legal metadata tracking or license compliance verification, requiring integration with external standards like SPDX 3.0 for comprehensive workflow analysis.
In parallel, recent knowledge graph-based efforts such as CRUX~\cite{wang2022crux} and ModsNet~\cite{wang2023selecting, wang2024modsnet} construct ML workflow knowledge bases to support workflow-centric queries, such as retrieving the best-performing models for a given dataset.
Other efforts, such as AiiDA~\cite{huber2020aiida} and ProvONE~\cite{cao2016provone}, focus on fine-grained provenance tracking in scientific workflows. 
They provide graph-based models to capture process and data lineage. Inspired by these approaches, our proposed ML workflow representation adopts their principles but extends them to support automated license compliance analysis by formally representing the reasoning logic of license rules and analyzing dependencies among ML assets.

\subsection{ML Model Licensing}
Today, we have many OSS licenses to accommodate diverse publishing scenarios~\cite{rosen2005open}. However, do they still function as intended in model publishing scenarios? The answer is no. While these licenses aim to govern the use and distribution of software, they lack definitions of ML concepts, which compromises their effectiveness (ref. Section~\ref{sec:license}).
Some model licenses (or agreements) are also emerging, such as Llama 2 and Gemma~\cite{gemma2024}. 
However, most of these licenses are specifically designed to govern certain models or their derivatives and are not as open as they claim to be~\cite{liesenfeld2024rethinking}. 
Meanwhile, Contractor \textit{et al.}~\cite{contractor2022behavioral} proposed OpenRAIL-M, a model license derived from Apache-2.0. 
While this license offers good clarity, it enforces use behavior restrictions that render it non-compliant with open-source licenses like GPL-3.0~\cite{greenbaum2016the}.
In addition, although OpenRAIL-M has many variations, its license terms are quite homogenized and lack the flexibility needed to accommodate different model publishing scenarios, such as non-commercial use, open sourcing, and restrictions on sharing outputs. 
Therefore, a significant number of developers have opted to publish their models using CC licenses, such as CC-BY-NC-4.0, as an alternative to prohibit commercial use of their models.
However, these licenses also face the issue of losing effectiveness in the context of model publishing.
Such unstandardized licensing practices can lead to increased compliance issues and pose potential legal hazards in ML projects~\cite{duan2024modelgo}. 
To promote standardized model publication, we propose a new set of model licenses that provide a wider range of licensing options.

\section{MG Analyzer}
\label{sec:analyzer}

\begin{figure*}[t]
    \centering
    \includegraphics[width=0.95\linewidth]{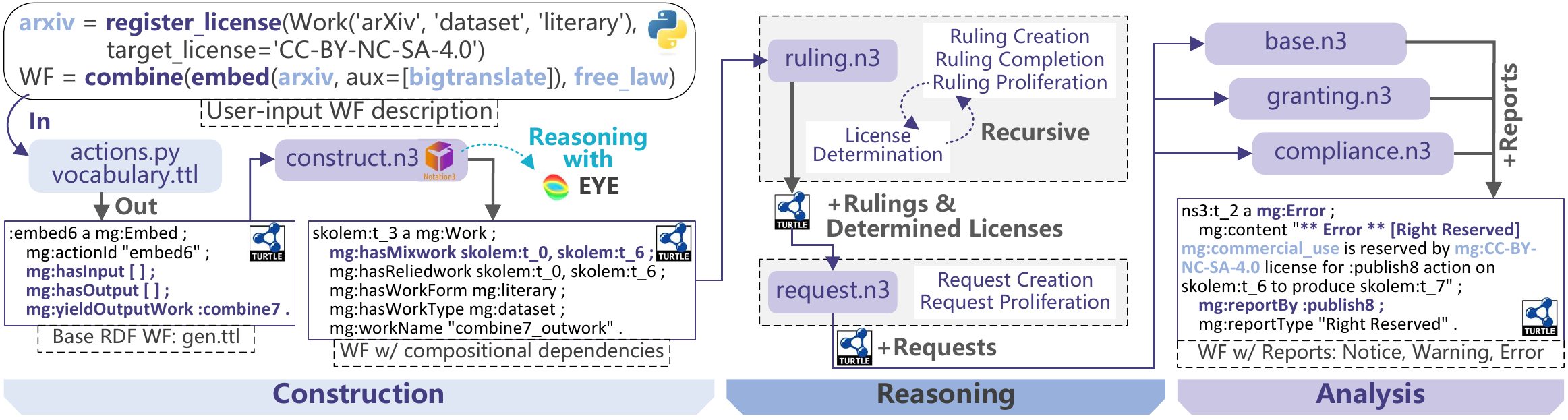}
    \caption{Overview of MG Analyzer ("mg" is the prefix for our proposed vocabulary).}
    \Description{}
    \label{fig:framework}
\end{figure*}

This section aims to introduce the specific design of MG Analyzer by exploring three questions: 
\emph{(i) How should we represent the workflows of ML projects to facilitate license-related noncompliance analysis? (ii) How do we establish a mapping from license text to reasonable rules? (iii) What types of license compliance issues can arise in ML projects, and how can we detect them?}
Before delving into the detailed design that answers these questions, we first provide an overview that serves as a roadmap for this section.
As illustrated in Figure~\ref{fig:framework}, the process of MG Analyzer is divided into three main parts: Construction, Reasoning, and Analysis.

In the \textbf{Construction Stage}, user-input workflow descriptions (written in Python) are converted into an RDF graph (saved as \emph{gen.ttl} in Turtle format) that contains the base information of the workflow. 
This conversion is achieved with the help of RDFLib~\cite{Krech_RDFLib_2023} and the \emph{MG Vocabulary}, which is part of our analyzer. 
RDFLib provides an API for writing RDF graphs, while the MG Vocabulary defines the specific semantics to represent the concepts and dependencies in ML projects.
Then, we apply reasoning rules (written in Notation3~\cite{arndt2023notation3}) for the complete workflow construction using the EYE reasoner~\cite{verborgh2015drawing}.
The reasoner concludes new properties that represent the input and output chains between components, followed by further reasoning to identify the \emph{compositional dependencies} among them (reflecting Question (i); see Section~\ref{sec:wf} for details).

The main tasks in \textbf{Reasoning Stage} involve concluding the \emph{definition dependencies} and \emph{rights-using} dependencies. 
This is achieved through two substeps.
First, a new property called \emph{ruling} is created to record the definition of the output work in relation to the input work within the context of licensing.
For example, if we merge GPL-licensed code into another software, the resulting work is considered a \emph{derivative} of the original work. 
This relationship, which we refer to as \emph{definition dependencies}, is crucial for determining the applicable license of the output work. 
We recursively identify such dependencies and ascertain the licenses of indeterminate works until all works in the workflow have a license.
Based on the RDF workflow graph with complete license assignments, we can execute the second step of reasoning, called \emph{request}, which infers \emph{rights-using} dependencies that represent the rights required for the work according to practical reuse methods (reflecting Question (ii); see Section~\ref{sec:rule} for details). 

So far, all necessary information for license analysis has been concluded before entering the final \textbf{Analysis Stage}. 
In this stage, MG Analyzer evaluates the validity of the base workflow information, checks for the satisfaction of rights granting, and assesses license compliance and conflicts. 
Entities of the class \emph{Report} are genreated to present these results in RDF format (reflecting Question (iii); see Section~\ref{sec:analysis} for details).

\subsection{ML Workflow Representation}
\label{sec:wf}
The representation of ML workflows, particularly when considering license analysis scenarios, differs significantly from common software workflows for the following three reasons.

\ding{172} ML workflows often involve various components (e.g., code, datasets, images, model weights, services), each governed by licenses from different frameworks. 
Additionally, non-standard licensing practices are prevalent in current ML projects~\cite{rajbahadur2021can, duan2024modelgo}, for instance, C4AI Command R+ model~\cite{cohereforai2024c4ai} is licensed under a free-content license: CC-BY-NC-4.0. 
Therefore, the representation should be flexible enough to cover such situations.

\ding{173} The component dependencies in ML workflows may be implicit and nested. 
For instance, Openjourney~\cite{openjourney2024prompthero} is fine-tuned based on StableDiffusion~\cite{rombach2022high} model and the data generated by Midjourney~\cite{midjourney2024terms}. 
In this case, knowledge from Midjourney is transferred to Openjourney without explicit compositional inclusion. 
Therefore, the representation should consider the multifarious dependencies present within ML projects.

\ding{174} The components' dependencies are also defined by the components' practical licenses and the ways they are reused. 
A common case in the OSS field is that republishing Software as a Service (SaaS) is considered to convey a \emph{derivative} under AGPL-3.0 but has \emph{no definition} under GPL-3.0. 
Therefore, terms like \emph{derivative} and \emph{independent} should be contextualized within specific licenses, and our representation should be capable of reflecting such meanings.

Therefore, we propose the MG Vocabulary to describe the properties and classes within ML workflows.
For the flexibility issue \ding{172}, we use the following terms to abstract key concepts of ML workflows:

\textbf{Work}: Represents the components (e.g., models, datasets), each with a unique Type and Form. 
A work can have a license assigned through a Register License Action, or its license can be determined through rules applied in the Reasoning Stage (ref Figure~\ref{fig:framework}).

\textbf{Action}: Represents operations performed on a Work, including Modify, Train and Combine, etc. 
In practice, we broaden the definition of these operations to make the vocabulary adaptable to different types of works. (See Table~\ref{tab:action} for more details.)

\textbf{Work Type}: Includes software, dataset, model, and mixed-type. It is used to describe both the nature of the work and to identify the types of materials intended by a license. 
We utilize this information to detect mismatches between a work's type and its license.

\textbf{Work Form}: Divided into three subclasses: Raw, Binary, and Service to provide flexibility. 
For example, source code, model weights, and corpus fall under Raw; compiled programs are considered Binary; and SaaS or online chat LLMs are categorized as Service.
Additionally, three general terms are offered: raw-form, binary-form, and service-form, which can work in conjunction with the work type. 
This approach helps represent concepts that lack a formal designation, such as "a dataset published as a service".
We use mixed-form to represent the cases involving collections of works.

\textbf{LicenseInfo}: This contains the essential license information derived from conditions, including the license name, ID, intended types of works, whether it is copyleft or permissive, as well as granted and reserved rights, etc.
While the license name is sufficient to describe the base workflow, to enable reasoning, LicenseInfo should bind rules, which we will discuss in the next section.

\begin{figure}[t]
    \centering
    \includegraphics[width=\linewidth]{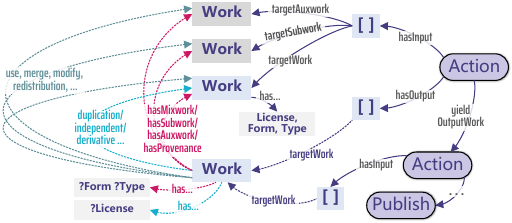}
    \caption{A Typical ML Workflow Represented by MG Analyzer. Dashed arrows with different colors indicate properties related to three types of dependencies: \textcolor{Construct}{compositional}, \textcolor{Ruling}{definition}, and \textcolor{Request}{rights-using} dependencies.}
    \Description{}
    \label{fig:wf}
\end{figure}

At this stage, we can describe a base ML workflow, as illustrated\footnote{For clarity, the “mg” prefix is omitted, and certain properties are merged or filtered.} schematically in Figure~\ref{fig:wf} (The RDF graph can be referred to in \emph{gen.ttl} in Figure~\ref{fig:framework}).
In this base workflow, the derived input and output of each Action can be represented as blank nodes, serving as placeholders. 
These placeholders will be populated by reasoning the output yielded by the previous action and the input for the current action, respectively.

For dependency issues discussed in \ding{173} and \ding{174}, we identify three potential types of dependencies in an ML project: \textbf{compositional, definition, and rights-using}. These dependencies are visually represented by different colored dashed arrows in Figure~\ref{fig:wf}.
\emph{Compositional dependencies} are categorized into four types: \emph{Mixwork}, \emph{Subwork}, \emph{Auxwork}, and \emph{Provenance}, each representing the containment relationships between input and output works.
For example, when the output work includes the input work or a part of it, the input is considered the Mixwork of the output. 
This type of dependency is vital for license analysis because all actions performed on the output work, including any rights usage, will proliferate to the Mixwork components.
For instance, when fine-tuning an ensemble model~\cite{furlanello2018born}, the fine-tuning operation will cascade to all submodels. 
Consequently, the license terms related to fine-tuning for each submodel are triggered, meaning that any constraints or rights imposed on the submodels must be honored. 
Additionally, if a work includes Mixworks in different forms, such as code and weights, we need to generalize the output's form to raw-form to accommodate these variations.
A similar approach applies to the work type as well.
Subwork and Auxwork are used to track works that are utilized by other works in the workflow, such as training datasets or distilled models. 
The key distinction is that Subwork is intended to be published alongside the output, which necessitates additional license analysis related to republishing.
Provenance is specifically used for the Register License action to indicate that the output is simply the input itself, bound to a license. In such cases, any further proliferation of dependencies should cease.

The \emph{definition dependencies} represent the relationships between works based on the definitions established by their licenses. 
For example, if a new work is created by modifying GPL-licensed code, that modification is considered a \emph{derivative} of the original work.
These dependencies should be understood in the context of the original license and can extend to subsequent actions if the same conditions are activated again (e.g., a \emph{derivative} of a \emph{derivative} is likely still a \emph{derivative}).
These dependencies are the main factor in determining the applicable license and restrictions for the output work. 
For example, if the original work is under GPL-3.0, the republication of \emph{derivatives} must also apply the GPL-3.0 license.
Additionally, a work may have multiple definition dependencies in a complex workflow, and these dependencies should be simultaneously satisfied during license determination (if possible; otherwise, an error should be reported).
The corresponding implementation in the MG Analyzer is illustrated in the \emph{Reasoning Stage} of Figure~\ref{fig:framework}. 
New instance nodes called \emph{Ruling} are created to track the definition dependencies and triggered rules for each work, determining their applicable licenses in an alternating manner.

The \emph{rights-using dependencies} describe the rights that must be granted for actions performed on works. 
For example, when executing a Train action on a model, it requires the rights to \emph{use} and \emph{modify} (termed as \emph{Usage} in MG Vocabulary) from the model's license\footnote{An example of such rights granting can be found in the Llama 2 Community License, which states, \emph{"You are granted a ... to use, reproduce, distribute, copy, create derivative works of, and make modifications to the Llama Materials."}.}.
Similarly, the rights-using dependencies should proliferate according to compositional dependencies.
For instance, the requirement to \emph{modify} a model extends to all its submodels.
In the MG Analyzer, we create new nodes called \emph{Request} to represent this dependency.
Additionally, it is insufficient to only check the granting rights; the reserved rights must also be verified. 
Depending on the clarity of the license text, some rights may either be explicitly granted or reserved.
Furthermore, certain license clauses can waive the requirement for specific rights. 
For instance, both GPL-3.0 and CC licenses include automatic relicensing clauses for downstream recipients, which eliminate the need for a \emph{sublicense} right.

By MG Vocabulary, we are able to describe complete ML workflows and represent the necessary dependencies for license analysis.
The \emph{compositional dependencies} are license-independent and can be reasoned from the base workflow. 
However, \emph{definition dependencies} and \emph{rights-using dependencies} are associated with specific rules expressed in natural language within each license.
To facilitate automated reasoning for these dependencies, the next step is to develop a viable method for encoding license terms into formal logic rules.

\subsection{License Rule Encoding and Reasoning}
\label{sec:rule}

Typically, licenses are designed to govern the use and distribution of specific types of works. 
For example, GPL-3.0 is tailored for source code and object code, while CC licenses focus on literary, musical, and artistic works. 
As a result, it is challenging to map their rules within a unified framework for logical reasoning.
Meanwhile, many ML projects actually incorporate non-standard licensing components, as mentioned in Section~\ref{sec:intro}.
If we consider these claimed licenses to be invalid, then they would not pose any license compliance issues.
However, the validity of these licenses depends on specific cases and the dispute resolution process by the jurisdictional courts in accordance with the applicable laws in different regions.
As a license analyzer, we aim to maximize the detection of all potential legal risks under various interpretations, rather than merely granting a green light with low confidence.
To this end, we perform three generalizations to encode license rules.

The first generalization is called \emph{fuzz form matching}. 
We broaden the definitions related to a work's form to encompass its general form.
For instance, we expand the license terms of GPL-3.0 concerning source code to include all forms of work in the Raw categories, such as model weights and corpus. 
In this way, we can extend the scope of interpretation of GPL-3.0 to cover models and datasets.

The second generalization is called \emph{composition-based rule alignment}, which aims to resolve the pervasive ambiguities across different licensing frameworks. 
This ambiguity often arises in non-standard scenarios, for example, when licensing a model under GPL-3.0, it may be unclear whether \emph{Model Aggregation} (a technology used in federated learning~\cite{li2021model}) triggers the "Aggregate" clause in GPL-3.0.
Therefore, we propose a composition-based method to align these rules. 
Specifically, we generalize the concept of action to represent the compositional relationships between input and output. 
For instance, the action \emph{Combine} signifies that the input work has been entirely included in the output work without modification. 
This action corresponds to terms such as "Link" and "Aggregate" in software licenses, "Collection" in CC licenses, and "MoE" in model licenses.
In the case of \emph{Model Aggregation}, which produces an output that contains parts of the input and is difficult to separate, it is not considered a \emph{Combine}. Consequently, according to our rule alignment, it will not activate the "Aggregate" clause in GPL-3.0.

The complete rule alignment method utilized in the MG Analyzer can be found in Table~\ref{tab:action}.
It is worth mentioning that the meanings of these actions have been broadened and may differ from their original definitions. 
In some cases, multiple actions may align with the same license terms. 
For instance, the license term "Modify" can align with both the actions \emph{Modify} and \emph{Amalgamate}, as such licenses do not distinguish the extent of changes made or whether those changes can be reverted.

\begin{figure}[t]
    \centering
    \includegraphics[width=0.85\linewidth]{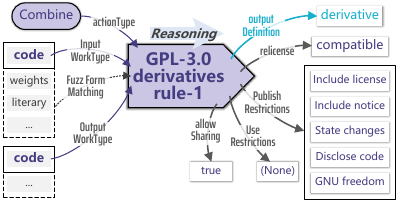}
    \caption{The Example of a Generalized GPL-3.0 Derivatives Rule in MG Analyzer.}
    \Description{}
    \label{fig:mgrule}
\end{figure}

The final aspect is \emph{applicable term generalization}, where we encode the triggering conditions of a license term into the following properties: range of input work forms, range of output work forms, and types of actions.
Figure~\ref{fig:mgrule} illustrates an example of the GPL-3.0 derivatives rule\footnote{The original GPL-3.0 license text reads: \emph{"You may convey a work based on the Program ... in the form of source code ... provided that you also meet all of these conditions: ... stating that you modified it ... it is released under this License ... keep intact all notices ... license the entire work, as a whole, under this License ..."} .}, which represents the Combine action applied to input works in code format, resulting in a code output.
This action triggers the "derivative" clause in GPL-3.0, indicating that the license of the output work must be compatible with GPL-3.0 (e.g., AGPL-3.0). Additionally, five restrictions must apply to the output work if it is to be republished, as dictated by this \emph{definition dependency}.
This rule does not include any \emph{Use Restrictions}, which apply to output works regardless of whether they are republished. 
We found this subtle distinction to be crucial in ML license analysis, as most OSS license terms are triggered by distribution, and their definitions of "distribution" typically exclude publishing as a service. 
However, in the case of models, the common deployment method is through a web interface,  leading to many OSS license restrictions being circumvented in such scenarios (ref Section~\ref{sec:case}).

Furthermore, multiple rules may lead to the same output definition, and we provide the option of \emph{fuzz form matching} by enabling \emph{fuzz rules} to enhance the interpretive capabilities of the MG Analyzer.
It is worth mentioning that the reasoning results behind the restrictions derived from these rules require further analysis for validation.
Taking the rule in Figure~\ref{fig:mgrule} as an example,  the presence of additional discrimination terms in the final work that violate GNU freedom~\cite{greenbaum2016the} dictates whether errors related to GNU freedom conflicts should be reported.
We present snippets of our encoded rules written in Turtle format in Appendix~\ref{apdx:rules}.
The list of supported licenses, whose terms have been encoded in MG Analyzer, is shown in Table~\ref{tab:licenses}, covering nearly all top-ranking licenses for published models on HuggingFace\footnote{\url{https://huggingface.co/models}}, a popular model publishing platform.

With MG Vocabulary, the ML workflow with compositional dependencies, and encoded license rules, we can reason to derive the definition and rights-using dependencies, thereby determining the applicable licenses for intermediate works. 
The license determination in the MG Analyzer follows an incremental and minimal noncompliance strategy, where the new license only applies to the incremental parts of the work without affecting the original work (a common practice in licensing). 
Furthermore, to avoid introducing additional compliance issues during analysis, we use \emph{Unlicense} as the default license when applicable.
However, an exception arises with the license proliferation clauses found in copyleft licenses, such as GPL-3.0 and OSL-3.0, which require that the entire new work be licensed under the same terms. 
Our analyzer incorporates reasoning logic to identify applicable licensing solutions (a snippet of logic can be found in Appendix~\ref{apdx:rules}), but unresolved conflicts may occur if multiple copyleft clauses are triggered. 
In such cases, the MG Analyzer will select one of these copyleft licenses and report an error during the \emph{Analysis Stage}.



\subsection{Compliance Analysis}
\label{sec:analysis}

\begin{table}[tp]
    \caption{List of \textcolor{Ruling}{Notices}, \textcolor{Construct}{Warnings}, and \textcolor{Request}{Errors} Reported by MG Analyzer. The triggered work is denoted as \emph{?work}. }
    \tiny
    \label{tab:report}
    \begin{tabular}{|c|p{1.9cm}|p{5cm}|}
    \hline
    \rowcolor[gray]{.8}
    Code & Report Type & Report Content \\ \hline
    
    N1 & \textcolor{Ruling}{Include License} & The original license file from \emph{?work} should be retained.  \\ \hline

    N2 & \textcolor{Ruling}{Include Notice} & The notices (e.g., attribution, copyright, patent, trademark) from \emph{?work} should be retained. \\ \hline

    N3 &\textcolor{Ruling}{State Changes} & A notice stating the modifications made to \emph{?work} should be provided. \\ \hline

    N4 &\textcolor{Ruling}{ImpACT Reports} & You need to complete a Derivative Impact Report. \\ \hline

    W1 &\textcolor{Construct}{License Type Mismatch} & Non-standard licensing of \emph{?work}. \\ \hline
    
    W2 &\textcolor{Construct}{Revocable License} & The license of \emph{?work} is revocable. \\ \hline
    
    W3 &\textcolor{Construct}{Possibly Revocable License} & The revocability of the license of \emph{?work} is not claimed. \\ \hline

    W4 &\textcolor{Construct}{Right Not Granted} & The required right is not explicitly granted by \emph{?work}. \\ \hline

    W5 &\textcolor{Construct}{Disclose Source Code} & This work should disclose its source code. \\ \hline

    W6 &\textcolor{Construct}{Disclose Unmodified Code} & The unmodified source code of \emph{?work} should be disclosed. \\ \hline

    W7 &\textcolor{Construct}{Use Behavior} & The use of this work must comply with the usage behavior restrictions of \emph{?work}. \\ \hline

    W8 &\textcolor{Construct}{Runtime Control} & There is a runtime restriction clause in \emph{?work} (e.g., forced updates). \\ \hline

    E1 &\textcolor{Request}{Wrong Work Type or Form} & The type of \emph{?work} is inconsistent with its form. \\ \hline

    E2 &\textcolor{Request}{Right Reserved} & The required right is reserved by the license of \emph{?work}. \\ \hline
    
    E3 &\textcolor{Request}{Not Allowed to Share} & Redistribution of this work is prohibited. \\ \hline

    E4 &\textcolor{Request}{Not Allowed to Sublicense} & Sublicensing of \emph{?work} is prohibited. \\ \hline

    E5 &\textcolor{Request}{Non-Commercial Use} & Commercial use of \emph{?work} is prohibited. \\ \hline

    E6 &\textcolor{Request}{Cannot Be Relicensed} & The license of this work is invalid because \emph{?work} cannot be relicensed, or relicensing is prohibited. \\ \hline

    E7 &\textcolor{Request}{GNU Freedom Conflict} & The additional terms applied in this work may violate the GNU freedom clauses of \emph{?work}. \\ \hline

    E8 &\textcolor{Request}{CC Freedom Conflict} & The additional terms applied in this work may violate the CC freedom clauses of \emph{?work}. \\ \hline

    E9 &\textcolor{Request}{Llama 2/3 Exclusive} & Using Llama 2/3's output in non-Llama 2/3 derivatives is prohibited. \\ \hline

    E10 &\textcolor{Request}{Exclusive License} & The additional terms applied in this work are prohibited by the license of \emph{?work}. \\ \hline

    \end{tabular}
\end{table}

At this stage, we establish all necessary dependency properties through automated reasoning to enable compliant license analysis. 
The analysis rules are designed to assess and report the validity of the base workflow information, the fulfillment of granted rights, work restrictions, and overall license compliance. 
In addition, the Publish action should be invoked to signify the completion of the workflow, along with an assigned public manner and work form. 
MG Analyzer considers three republication scenarios: internal, share, and sell, each of which typically involves different terms and conditions in the licenses.
For instance, if we publish the final work for sale, the related licenses should grant rights for redistribution, sublicensing, and commercial use. 

Appendix~\ref{apdx:rules} presents the logic code for rights granting analysis, and the full list of reported notices, warnings, and errors is shown in Table~\ref{tab:report}.
Due to its staged logic rule reasoning design, MG Analyzer has considerable extensibility, enabling the incorporation of additional licenses and analysis targets, provided that it can reason based on our proposed dependencies information.
Beyond compliance analysis, our vision is to promote ML workflow supply chain management and improve FAIRness~\cite{wilkinson2016fair} in model publishing. 
We position our vocabulary and tool as a first step toward Linked Open Model Production Data~\cite{bizer2008linked}.

\section{MG Licenses and Comparison Results}
\label{sec:license}

\begin{table*}[tp]
    \caption{List of MG Analyzer-Supported Licenses \& Agreements (including \href{https://github.com/morningD/ModelGo-Analyzer/tree/main/MGLicense}{\textcolor{Construct}{MG Licenses V1.0}}) with Their Comparisons in Clarity and Freedom. Grouped by OSS, Free-Content (\&Dataset), Model and sorted first by clarity score, then by freedom score.}
    \tiny
    \label{tab:licenses}
    \begin{tabular}{|lcccccccccccccccc|}
        \hline
        \multicolumn{1}{|l|}{\multirow{2}{*}{License Name}} & \multicolumn{4}{c|}{Clarity of Definitions} & \multicolumn{3}{c|}{Freedom of Verbatim Copy} & \multicolumn{5}{c|}{Freedom of Derivative} & \multicolumn{2}{c|}{Freedom of Use} & \multicolumn{1}{c|}{\multirow{2}{*}{Clarity}} & \multirow{2}{*}{Freedom} \\ \cline{2-15}

        \multicolumn{1}{|l|}{} & \multicolumn{1}{l|}{Prefixes} & \multicolumn{1}{l|}{Rights} & \multicolumn{1}{l|}{Rules} & \multicolumn{1}{l|}{Remote} & \multicolumn{1}{l|}{Share} & \multicolumn{1}{l|}{Close} & \multicolumn{1}{l|}{Non-excl.} & \multicolumn{1}{l|}{Share} & \multicolumn{1}{l|}{Close} & \multicolumn{1}{l|}{Non-excl.} & \multicolumn{1}{l|}{Sublicense} & \multicolumn{1}{l|}{Attribute} & \multicolumn{1}{l|}{Comm.} & \multicolumn{1}{l|}{Behav.} & \multicolumn{1}{l|}{} & \\ 
        
        \hline 
        
        \multicolumn{1}{|r|}{AGPL-3.0} & \cmark & \cmark & \pmark & \multicolumn{1}{c|}{\cmark} & \pmark & \xmark & \multicolumn{1}{c|}{\pmark} & \pmark & \xmark & \pmark & \pmark & \multicolumn{1}{c|}{\xmark} & \cmark & \multicolumn{1}{c|}{\cmark} & \multicolumn{1}{c|}{3.5} & 4.5 \\
        
        \multicolumn{1}{|r|}{AFL-3.0} & \pmark & \cmark & \pmark & \multicolumn{1}{c|}{\cmark} & \pmark & \cmark & \multicolumn{1}{c|}{\cmark} & \pmark & \pmark & \cmark & \cmark & \multicolumn{1}{c|}{\xmark} & \cmark & \multicolumn{1}{c|}{\cmark} & \multicolumn{1}{c|}{3.0} & 7.5 \\

        \multicolumn{1}{|r|}{OSL-3.0} & \pmark & \cmark & \pmark & \multicolumn{1}{c|}{\cmark} & \pmark & \cmark & \multicolumn{1}{c|}{\xmark} & \pmark & \xmark & \xmark & \cmark & \multicolumn{1}{c|}{\xmark} & \cmark & \multicolumn{1}{c|}{\cmark} & \multicolumn{1}{c|}{3.0} & 5 \\

        \multicolumn{1}{|r|}{Apache-2.0} & \cmark & \cmark & \pmark & \multicolumn{1}{c|}{\xmark} & \pmark & \cmark & \multicolumn{1}{c|}{\cmark} & \pmark & \pmark &\cmark & \cmark & \multicolumn{1}{c|}{\xmark} & \cmark & \multicolumn{1}{c|}{\cmark} & \multicolumn{1}{c|}{2.5} & 7.5 \\

        \multicolumn{1}{|r|}{LGPL-3.0} & \cmark & \cmark & \pmark & \multicolumn{1}{c|}{\xmark} & \pmark & \xmark & \multicolumn{1}{c|}{\cmark} & \pmark & \cmark & \cmark & \pmark & \multicolumn{1}{c|}{\cmark} & \cmark & \multicolumn{1}{c|}{\cmark} & \multicolumn{1}{c|}{2.5} & 7.5 \\

        \multicolumn{1}{|r|}{Artistic-2.0} & \pmark & \pmark & \pmark & \multicolumn{1}{c|}{\cmark} & \pmark & \xmark & \multicolumn{1}{c|}{\cmark} & \pmark & \xmark & \cmark & \cmark & \multicolumn{1}{c|}{\xmark} & \cmark & \multicolumn{1}{c|}{\cmark} & \multicolumn{1}{c|}{2.5} & 6.0 \\

        \multicolumn{1}{|r|}{GPL-3.0} & \cmark & \cmark & \pmark & \multicolumn{1}{c|}{\xmark} & \pmark & \xmark & \multicolumn{1}{c|}{\pmark} & \pmark & \xmark & \pmark & \pmark & \multicolumn{1}{c|}{\xmark} & \cmark & \multicolumn{1}{c|}{\cmark} & \multicolumn{1}{c|}{2.5} & 4.5 \\
        
        \multicolumn{1}{|r|}{ECL-2.0} & \pmark & \cmark & \pmark & \multicolumn{1}{c|}{\xmark} & \pmark & \cmark & \multicolumn{1}{c|}{\cmark} & \pmark & \pmark & \cmark & \cmark & \multicolumn{1}{c|}{\xmark} & \cmark & \multicolumn{1}{c|}{\cmark} & \multicolumn{1}{c|}{2.0} & 7.5 \\ 

        \multicolumn{1}{|r|}{Unlicense} & \pmark & \pmark & \pmark & \multicolumn{1}{c|}{\xmark} & \cmark & \cmark & \multicolumn{1}{c|}{\cmark} & \cmark & \cmark & \cmark & \cmark & \multicolumn{1}{c|}{\cmark} & \cmark & \multicolumn{1}{c|}{\cmark} & \multicolumn{1}{c|}{1.5} & 10 \\

        \multicolumn{1}{|r|}{MIT} & \pmark & \pmark & \pmark & \multicolumn{1}{c|}{\xmark} & \pmark & \cmark & \multicolumn{1}{c|}{\cmark} & \cmark & \cmark & \cmark & \cmark & \multicolumn{1}{c|}{\xmark} & \cmark & \multicolumn{1}{c|}{\cmark} & \multicolumn{1}{c|}{1.5} & 8.5 \\

        \multicolumn{1}{|r|}{GPL-2.0} & \xmark & \cmark & \pmark & \multicolumn{1}{c|}{\xmark} & \pmark & \xmark & \multicolumn{1}{c|}{\cmark} & \pmark & \xmark & \cmark & \pmark & \multicolumn{1}{c|}{\xmark} & \cmark & \multicolumn{1}{c|}{\cmark} & \multicolumn{1}{c|}{1.5} & 5.5 \\

        \multicolumn{1}{|r|}{LGPL-2.1} & \xmark & \cmark & \pmark & \multicolumn{1}{c|}{\xmark} & \pmark & \xmark & \multicolumn{1}{c|}{\cmark} & \pmark & \xmark & \cmark & \pmark & \multicolumn{1}{c|}{\xmark} & \cmark & \multicolumn{1}{c|}{\cmark} & \multicolumn{1}{c|}{1.5} & 5.5 \\
        
        \multicolumn{1}{|r|}{BSD-3-Clause} & \xmark & \pmark & \pmark & \multicolumn{1}{c|}{\xmark} & \pmark & \cmark & \multicolumn{1}{c|}{\cmark} & \pmark & \cmark & \cmark & \cmark & \multicolumn{1}{c|}{\xmark} & \pmark & \multicolumn{1}{c|}{\cmark} & \multicolumn{1}{c|}{1.0} & 7.5 \\

        \multicolumn{1}{|r|}{BSD-3-Clause-Clear} & \xmark & \pmark & \pmark & \multicolumn{1}{c|}{\xmark} & \pmark & \cmark & \multicolumn{1}{c|}{\cmark} & \pmark & \cmark & \cmark & \cmark & \multicolumn{1}{c|}{\xmark} & \pmark & \multicolumn{1}{c|}{\cmark} & \multicolumn{1}{c|}{1.0} & 7.5 \\

        \multicolumn{1}{|r|}{BSD-2-Clause} & \xmark & \pmark & \pmark & \multicolumn{1}{c|}{\xmark} & \pmark & \cmark & \multicolumn{1}{c|}{\cmark} & \pmark & \cmark & \cmark & \cmark & \multicolumn{1}{c|}{\xmark} & \pmark & \multicolumn{1}{c|}{\cmark} & \multicolumn{1}{c|}{1.0} & 7.5 \\

        \multicolumn{1}{|r|}{WTFPL-2.0} & \xmark & \xmark & \xmark & \multicolumn{1}{c|}{\xmark} & \cmark & \cmark & \multicolumn{1}{c|}{\cmark} & \cmark & \cmark & \cmark & \cmark & \multicolumn{1}{c|}{\cmark} & \cmark & \multicolumn{1}{c|}{\cmark} & \multicolumn{1}{c|}{0} & 10 \\

        \hline 

        \multicolumn{1}{|r|}{CC0-1.0} & \cmark & \cmark & \cmark & \multicolumn{1}{c|}{\xmark} & \cmark & \cmark & \multicolumn{1}{c|}{\cmark} & \cmark & \cmark & \cmark & \cmark * & \multicolumn{1}{c|}{\cmark} & \cmark & \multicolumn{1}{c|}{\cmark} & \multicolumn{1}{c|}{3.0} & 10 \\

        \multicolumn{1}{|r|}{ODC-By-1.0} & \pmark & \pmark & \cmark & \multicolumn{1}{c|}{\cmark} & \pmark & \cmark & \multicolumn{1}{c|}{\cmark} & \pmark & \cmark & \cmark & \pmark & \multicolumn{1}{c|}{\xmark} & \cmark & \multicolumn{1}{c|}{\cmark} & \multicolumn{1}{c|}{3.0} & 7.5 \\

        \multicolumn{1}{|r|}{PDDL-1.0} & \cmark & \pmark & \cmark & \multicolumn{1}{c|}{\xmark} & \cmark & \cmark & \multicolumn{1}{c|}{\cmark} & \cmark & \cmark & \cmark & \cmark & \multicolumn{1}{c|}{\cmark} & \cmark & \multicolumn{1}{c|}{\cmark} & \multicolumn{1}{c|}{2.5} & 10 \\

        \multicolumn{1}{|r|}{CC-BY-4.0} & \cmark & \cmark & \pmark & \multicolumn{1}{c|}{\xmark} & \pmark & \cmark & \multicolumn{1}{c|}{\pmark} & \pmark & \pmark & \pmark & \pmark & \multicolumn{1}{c|}{\xmark} & \cmark & \multicolumn{1}{c|}{\cmark} & \multicolumn{1}{c|}{2.5} & 6.0 \\

        \multicolumn{1}{|r|}{CC-BY-SA-4.0} & \cmark & \cmark & \pmark & \multicolumn{1}{c|}{\xmark} & \pmark & \cmark & \multicolumn{1}{c|}{\pmark} & \pmark & \pmark & \pmark & \pmark & \multicolumn{1}{c|}{\xmark} & \cmark & \multicolumn{1}{c|}{\cmark} & \multicolumn{1}{c|}{2.5} & 6.0 \\

        \multicolumn{1}{|r|}{CC-BY-NC-4.0} & \cmark & \cmark & \pmark & \multicolumn{1}{c|}{\xmark} & \pmark & \cmark & \multicolumn{1}{c|}{\pmark} & \pmark & \pmark & \pmark & \pmark & \multicolumn{1}{c|}{\xmark} & \xmark & \multicolumn{1}{c|}{\cmark} & \multicolumn{1}{c|}{2.5} & 5.0 \\

        \multicolumn{1}{|r|}{CC-BY-NC-SA-4.0} & \cmark & \cmark & \pmark & \multicolumn{1}{c|}{\xmark} & \pmark & \cmark & \multicolumn{1}{c|}{\pmark} & \pmark & \pmark & \pmark & \pmark & \multicolumn{1}{c|}{\xmark} & \xmark & \multicolumn{1}{c|}{\cmark} & \multicolumn{1}{c|}{2.5} & 5.0 \\

        \multicolumn{1}{|r|}{CC-BY-ND-4.0} & \cmark & \cmark & \pmark & \multicolumn{1}{c|}{\xmark} & \pmark & \cmark & \multicolumn{1}{c|}{\pmark} & \xmark & \namark ** & \namark & \namark & \multicolumn{1}{c|}{\namark} & \cmark & \multicolumn{1}{c|}{\cmark} & \multicolumn{1}{c|}{2.5} & 4.0 \\

        \multicolumn{1}{|r|}{CC-BY-NC-ND-4.0} & \cmark & \cmark & \pmark & \multicolumn{1}{c|}{\xmark} & \pmark & \cmark & \multicolumn{1}{c|}{\pmark} & \xmark & \namark ** & \namark & \namark & \multicolumn{1}{c|}{\namark} & \xmark & \multicolumn{1}{c|}{\cmark} & \multicolumn{1}{c|}{2.5} & 3.0 \\

        \multicolumn{1}{|r|}{GFDL} & \cmark & \pmark & \pmark & \multicolumn{1}{c|}{\xmark} & \pmark & \xmark & \multicolumn{1}{c|}{\xmark} & \pmark & \xmark & \xmark & \pmark & \multicolumn{1}{c|}{\xmark} & \cmark & \multicolumn{1}{c|}{\cmark} & \multicolumn{1}{c|}{2.0} & 3.5 \\

        \multicolumn{1}{|r|}{C-UDA} & \cmark & \xmark & \pmark & \multicolumn{1}{c|}{\xmark} & \cmark & \cmark & \multicolumn{1}{c|}{\cmark} & \pmark & \cmark & \cmark & \xmark & \multicolumn{1}{c|}{\xmark} & \xmark & \multicolumn{1}{c|}{\xmark} & \multicolumn{1}{c|}{1.5} & 5.5 \\

        \multicolumn{1}{|r|}{LGPLLR} & \xmark & \cmark & \pmark & \multicolumn{1}{c|}{\xmark} & \pmark & \cmark & \multicolumn{1}{c|}{\xmark} & \pmark & \pmark & \xmark & \pmark & \multicolumn{1}{c|}{\xmark} & \pmark & \multicolumn{1}{c|}{\cmark} & \multicolumn{1}{c|}{1.5} & 4.5 \\

        \hline 

        \multicolumn{1}{|r|}{\textcolor{Construct}{\textbf{MG0}}} & \cmark & \cmark & \cmark & \multicolumn{1}{c|}{\cmark} & \pmark & \cmark & \multicolumn{1}{c|}{\cmark} & \pmark & \cmark & \cmark & \cmark & \multicolumn{1}{c|}{\pmark} & \cmark & \multicolumn{1}{c|}{\cmark} & \multicolumn{1}{c|}{4.0} & 8.5 \\

        \multicolumn{1}{|r|}{\textcolor{Construct}{\textbf{MG-BY}}} & \cmark & \cmark & \cmark & \multicolumn{1}{c|}{\cmark} & \pmark & \cmark & \multicolumn{1}{c|}{\cmark} & \pmark & \pmark & \cmark & \cmark & \multicolumn{1}{c|}{\xmark} & \cmark & \multicolumn{1}{c|}{\cmark} & \multicolumn{1}{c|}{4.0} & 7.5 \\

        \multicolumn{1}{|r|}{\textcolor{Construct}{\textbf{MG-BY-RAI}}} & \cmark & \cmark & \cmark & \multicolumn{1}{c|}{\cmark} & \pmark & \cmark & \multicolumn{1}{c|}{\cmark} & \pmark & \pmark & \cmark & \cmark & \multicolumn{1}{c|}{\xmark} & \cmark & \multicolumn{1}{c|}{\xmark} & \multicolumn{1}{c|}{4.0} & 6.5 \\

        \multicolumn{1}{|r|}{{\ddag}~OpenRAIL-M} & \cmark & \cmark & \cmark & \multicolumn{1}{c|}{\cmark} & \pmark & \cmark & \multicolumn{1}{c|}{\cmark} & \pmark & \pmark & \cmark & \cmark & \multicolumn{1}{c|}{\xmark} & \cmark & \multicolumn{1}{c|}{\xmark} & \multicolumn{1}{c|}{4.0} & 6.5 \\

        \multicolumn{1}{|r|}{\textcolor{Construct}{\textbf{MG-BY-NC}}} & \cmark & \cmark & \cmark & \multicolumn{1}{c|}{\cmark} & \pmark & \cmark & \multicolumn{1}{c|}{\cmark} & \pmark & \pmark & \cmark & \pmark & \multicolumn{1}{c|}{\xmark} & \xmark & \multicolumn{1}{c|}{\cmark} & \multicolumn{1}{c|}{4.0} & 6.0 \\

        \multicolumn{1}{|r|}{\textcolor{Construct}{\textbf{MG-BY-OS}}} & \cmark & \cmark & \cmark & \multicolumn{1}{c|}{\cmark} & \pmark & \xmark & \multicolumn{1}{c|}{\cmark} & \pmark & \xmark & \cmark & \cmark & \multicolumn{1}{c|}{\xmark} & \cmark & \multicolumn{1}{c|}{\cmark} & \multicolumn{1}{c|}{4.0} & 6.0 \\

        \multicolumn{1}{|r|}{\textcolor{Construct}{\textbf{MG-BY-NC-RAI}}} & \cmark & \cmark & \cmark & \multicolumn{1}{c|}{\cmark} & \pmark & \cmark & \multicolumn{1}{c|}{\cmark} & \pmark & \pmark & \cmark & \pmark & \multicolumn{1}{c|}{\xmark} & \xmark & \multicolumn{1}{c|}{\xmark} & \multicolumn{1}{c|}{4.0} & 5.0 \\

        \multicolumn{1}{|r|}{\textcolor{Construct}{\textbf{MG-BY-NC-OS}}} & \cmark & \cmark & \cmark & \multicolumn{1}{c|}{\cmark} & \pmark & \xmark & \multicolumn{1}{c|}{\cmark} & \pmark & \xmark & \cmark & \pmark & \multicolumn{1}{c|}{\xmark} & \xmark & \multicolumn{1}{c|}{\cmark} & \multicolumn{1}{c|}{4.0} & 4.5 \\

        \multicolumn{1}{|r|}{\textcolor{Construct}{\textbf{MG-BY-ND}}} & \cmark & \cmark & \cmark & \multicolumn{1}{c|}{\cmark} & \pmark & \cmark & \multicolumn{1}{c|}{\cmark} & \xmark & \namark & \namark & \namark & \multicolumn{1}{c|}{\namark} & \cmark & \multicolumn{1}{c|}{\cmark} & \multicolumn{1}{c|}{4.0} & 4.5 \\

        \multicolumn{1}{|r|}{\textcolor{Construct}{\textbf{MG-BY-NC-ND}}} & \cmark & \cmark & \cmark & \multicolumn{1}{c|}{\cmark} & \pmark & \cmark & \multicolumn{1}{c|}{\cmark} & \xmark & \namark & \namark & \namark & \multicolumn{1}{c|}{\namark} & \xmark & \multicolumn{1}{c|}{\cmark} & \multicolumn{1}{c|}{4.0} & 3.5 \\

        \multicolumn{1}{|r|}{{\dag}~OPT-175B} & \cmark & \pmark & \cmark & \multicolumn{1}{c|}{\cmark} & \pmark & \cmark & \multicolumn{1}{c|}{\cmark} & \pmark & \cmark & \cmark & \xmark & \multicolumn{1}{c|}{\xmark} & \xmark & \multicolumn{1}{c|}{\xmark} & \multicolumn{1}{c|}{3.5} & 5.0 \\

        \multicolumn{1}{|r|}{{\dag}~Llama3} & \pmark & \xmark & \cmark & \multicolumn{1}{c|}{\cmark} & \pmark & \cmark & \multicolumn{1}{c|}{\cmark} & \pmark & \cmark & \cmark & \xmark & \multicolumn{1}{c|}{\xmark} & \pmark & \multicolumn{1}{c|}{\xmark} & \multicolumn{1}{c|}{2.5} & 5.5 \\

        \multicolumn{1}{|r|}{{\dag}~Llama3.1} & \pmark & \xmark & \cmark & \multicolumn{1}{c|}{\cmark} & \pmark & \cmark & \multicolumn{1}{c|}{\cmark} & \pmark & \cmark & \cmark & \xmark & \multicolumn{1}{c|}{\xmark} & \pmark & \multicolumn{1}{c|}{\xmark} & \multicolumn{1}{c|}{2.5} & 5.5 \\

        \multicolumn{1}{|r|}{{\dag}~{\textbullet}~AI2-ImpACT-LR} & \pmark & \xmark & \cmark & \multicolumn{1}{c|}{\cmark} & \pmark & \cmark & \multicolumn{1}{c|}{\cmark} & \pmark & \cmark & \cmark & \xmark & \multicolumn{1}{c|}{\xmark} & \pmark & \multicolumn{1}{c|}{\xmark} & \multicolumn{1}{c|}{2.5} & 5.5 \\

        \multicolumn{1}{|r|}{{\dag}~{\textbullet}~AI2-ImpACT-MR} & \pmark & \xmark & \cmark & \multicolumn{1}{c|}{\cmark} & \xmark & \namark & \multicolumn{1}{c|}{\namark} & \pmark & \cmark & \cmark & \xmark & \multicolumn{1}{c|}{\xmark} & \xmark & \multicolumn{1}{c|}{\xmark} & \multicolumn{1}{c|}{2.5} & 2.0 \\

        \multicolumn{1}{|r|}{{\dag}~{\textbullet}~AI2-ImpACT-HR} & \pmark & \xmark & \cmark & \multicolumn{1}{c|}{\cmark} & \xmark & \namark & \multicolumn{1}{c|}{\namark} & \xmark & \namark & \namark & \namark & \multicolumn{1}{c|}{\namark} & \xmark & \multicolumn{1}{c|}{\xmark} & \multicolumn{1}{c|}{2.5} & 0 \\

        \multicolumn{1}{|r|}{{\dag}~Gemma} & \xmark & \xmark & \cmark & \multicolumn{1}{c|}{\cmark} & \pmark & \cmark & \multicolumn{1}{c|}{\cmark} & \pmark & \pmark & \cmark & \xmark & \multicolumn{1}{c|}{\xmark} & \pmark & \multicolumn{1}{c|}{\xmark} & \multicolumn{1}{c|}{2.0} & 4.5 \\

        \multicolumn{1}{|r|}{{\dag}~Llama2} & \pmark & \xmark & \cmark & \multicolumn{1}{c|}{\xmark} & \pmark & \cmark & \multicolumn{1}{c|}{\cmark} & \pmark & \cmark & \cmark & \xmark & \multicolumn{1}{c|}{\xmark} & \pmark & \multicolumn{1}{c|}{\xmark} & \multicolumn{1}{c|}{1.5} & 5.5 \\

        \hline\hline
        \multicolumn{17}{|p{16cm}|}{
            \textbf{Header Definitions:} \newline
            \textbf{Prefixes}: \ding{51} The license explicitly includes sufficient prefixes that clearly describe scope and conditions of granting rights (e.g., revocable, sublicensable); $\bm{\approx}$ Some important prefixes are indeterminate; \ding{55} No prefixes are declared. \newline
            \textbf{Rights}: \ding{51} The license explicitly declares whether a patent license or a copyright license is granted; $\bm{\approx}$ Only the granting of a patent license or copyright license is stated; \ding{55} No explicit grant of either is provided. \newline
            \textbf{Rules}: \ding{51} The license terms cover all actions listed in Table~\ref{tab:action}; $\bm{\approx}$ Some actions fall outside the definition of this license; \ding{55} Almost no rules are set forth. \newline
            \textbf{Remote}: \ding{51} The license considers remote access situations (e.g., via API, Web, SaaS); \ding{55} No definitions or rules regarding remote access behaviors are set forth.\newline
            \textbf{Share}: \ding{51} The license permits the sharing of verbatim copies/derivatives created by you without any restrictions; $\bm{\approx}$ Some restrictions apply to sharing; \ding{55} Sharing verbatim copies/derivatives is prohibited. \newline
            \textbf{Close}: \ding{51} The license does not require you to disclose the source files of verbatim copies/derivatives created by you; $\bm{\approx}$ Modification statements are required; \ding{55} You must disclose the source files of your created copies/derivatives. \newline
            \textbf{Non-excl}usive: \ding{51} The license does not restrict you from adding new terms when republishing; $\bm{\approx}$ Certain types of terms are prohibited in republishing; \ding{55} All republishing must adhere to the original terms and conditions. \newline
            \textbf{Sublicense}: \ding{51} The license explicitly grants sublicensing rights; $\bm{\approx}$ The license prohibits sublicensing but offers automatic licensing instead; \ding{55} Sublicensing is either prohibited or not explicitly permitted. \newline
            \textbf{Attribute}: \ding{51} The license does not require retaining the original attribution and licenses in redistributed derivatives; $\bm{\approx}$ Attribution or license must be retained; \ding{55} Redistributed derivatives must retain the  attributions and licenses. \newline
            \textbf{Comm}ercial: \ding{51} The license explicitly grants commercial rights; $\bm{\approx}$ Commercial rights are not explicitly granted but not reserved either, or compromised commercial rights are granted; \ding{55} Commercial rights are reserved. \newline
            \textbf{Behav}ioral: \ding{51} The license does not restrict user behaviors; $\bm{\approx}$ Includes runtime controls (e.g., forced updates); \ding{55} Certain behaviors involving the licensed materials or derivatives are prohibited (e.g., harming, medical advice). \newline          
            \textbf{Clarity}/\textbf{Freedom} Score: \ding{51} $+1.0$, $\bm{\approx}$ $+0.5$, \ding{55} $+0$, n/a: $+0$ . Maximum Clarity Score: 4.0, Maximum Freedom Score: 10. \newline
            \textbf{Explanations:} \newline
            *~Although CC0-1.0 explicitly states that sublicensing is not allowed, sublicensing becomes unnecessary due to the Waiver of Rights. \newline
            **~Since CC-BY-ND-4.0 and CC-BY-NC-ND-4.0 prohibit the sharing of derivatives, judgments regarding redistributed derivatives are marked as "n/a" in the table. \newline
            \dag~These licenses (or terms of use, or agreements) are specifically drafted for certain products and are not intended for general model publishing purposes. \newline
            \ddag~As there are no fundamental differences between CreativeML Open RAIL-M, OpenRAIL++-M, BigCode Open RAIL-M, BigScience RAIL, and BigScience Open RAIL-M, these licenses are grouped under OpenRAIL-M. \newline
            \textbullet~We have used an archive of the AI2 ImpaACT license; the version is 2.0, with an effective date of January 8, 2024.
        } \\ \hline
    \end{tabular}
\end{table*}

Although the MG Analyzer can identify potential non-compliance in existing ML projects, it does not offer an effective solution to prevent such issues in the future. 
To address this, we conducted a survey of the most widely used licenses for models published on HuggingFace and identified three major causes of license noncompliance in current ML projects: non-standard licensing, lack of general model licenses, and insufficiently defined licenses. 
The statistical results of a previous study~\cite{duan2024modelgo} support part of our findings.

To reveal the underlying dilemmas in model licensing, we provide comprehensive comparisons of these licenses in Table~\ref{tab:licenses}.
Based on the terms of these licenses, we evaluated each license's clarity score and freedom score, each encompassing sub-items as defined in the table.
A higher clarity score indicates that the license is more clearly defined in model publishing scenarios, while a higher freedom\footnote{Our freedom score reflects only the amount of restrictions stipulated in a license and should not be confused with the definitions of "Freedom" in free software~\cite{perens1999open}.} score signifies fewer restrictions on republished copies and derivatives.
The significant findings are summarized below:

\ding{172} \textbf{For OSS licenses}, most are not well defined in the context of model publishing, primarily due to the absence of clauses addressing ML activities (Rules) and the lack of coverage for publishing as a service (Remote).
This implies that mainstream model deployment practices, which often provide models as web services, are likely to circumvent the governance of these licenses. 
Additionally, commercial use and behavioral restrictions are not stipulated in most OSS licenses, which are common requirements in model publishing.

\ding{173} \textbf{For free-content and dataset licenses}, their average clarity score is only slightly better than that of OSS licenses, and they also lack coverage for publishing as a service.
However, some CC licenses, such as CC-BY-NC-4.0, offer additional options that prohibit the commercial use of the work, unlike OSS licenses. 
This may explain why many models are published under these licenses\footnote{There are 9,730 models on HuggingFace licensed under CC-BY-NC-4.0 (more than Llama 2), and 770 of them have garnered 1k+ downloads (Accessed on May 29. 2025).}, despite the fact that they were not originally drafted for models. 

\ding{174} \textbf{For model licenses}, aside from the proposed MG licenses, most model licenses are not intended for general publishing purposes. 
For example, the terms in Llama 2 license is specifically drafted to govern the Llama 2 model and its derivatives\footnote{For example, Clause 1.v. of the Llama 2 license reads: \emph{"You will not use the Llama Materials or any output or results of the Llama Materials to improve any other large language model (excluding Llama 2 or derivative works thereof)."}.}, failing to meet reusable standards.
Additionally, since the purpose of these licenses is often to protect the IP rights of proprietary models, they are usually revocable and prohibit sublicensing.
Although a set of well-defined licenses known as OpenRAIL-M~\cite{contractor2022behavioral} exists, their nearly identical rules make it difficult to accommodate the diverse needs in model publishing.

Based on the above findings, we conclude that it is necessary to draft new standardized and flexible licenses for general model publishing.
To address this, we collaborated with a law firm to draft a set of model licenses, tentatively referred to as MG Licenses. 
As reflecting in Table~\ref{tab:licenses}, our licenses include specific definitions and terms related to ML concepts, offering greater clarity compared to OSS and free-content licenses. 
Most importantly, following the philosophy of CC licenses, MG Licenses are designed to be flexible and easy to use, offering five options to accommodate various publishing scenarios.
The options include: BY (Attribution), NC (Non-Commercial), ND (No Derivatives), RAI (Responsible Use of AI), and OS (Open Source).
We drafted nine preset licenses using these options, including MG0, which does not apply any options.
The flexibility of our licenses is reflected in Table~\ref{tab:licenses}, where their freedom scores are evenly distributed between 3.5 and 8.5, indicating that they form a superset of all other model licenses\footnote{AI2-ImpACT-MR and AI2-ImpACT-HR prohibit sharing copies, we do not consider such restrictions to be common in model publishing, so we have excluded them. For details on how to substitute other licenses with MG licenses, refer to Appendix~\ref{apdx:model_sheet}.}.
To promote transparency, we introduced a \emph{Model Sheet} as an attachment to each MG License, inspired by MDL~\cite{benjamin2019towards}. 
This sheet assists model users in understanding the rights and restrictions granted by the license terms and helps model developers identify the most suitable license for their needs (see Attachment A of \href{https://github.com/morningD/ModelGo-Analyzer/blob/main/MGLicense/MG0.txt}{MG0-1.0}).

In the next section, we examine which licenses protect models from misuse beyond their intended purposes (e.g., allowing closed-sourcing under GPL). We evaluate popular model licenses and the MG Licenses using MG Analyzer.



\section{Preliminary License Analysis Results}
\label{sec:case}
\begin{figure}[t]
    \centering
    \includegraphics[width=.95\linewidth]{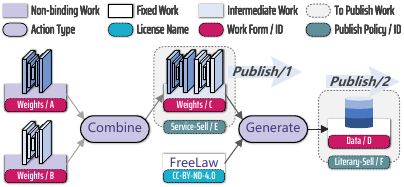}
    \caption{Example Workflow: Combine Models, Then Publish.}
    \Description{}
    \label{fig:case1}
\end{figure}

This section seeks to answer a key question: \emph{Should I continue using traditional OSS and free-content licenses to publish my model, and what are the associated risks?}
We evaluate commonly adopted licenses using MG Analyzer to assess their effectiveness in this context.
The example workflow involves combining two models and publishing them as a service, as shown in Figure~\ref{fig:case1}.
Here, we consider two scenarios: 1) publishing the combined model as a service; 2) publishing the data generated from the combined model.
Models A and B are non-binding works that correspond to the settings in Table~\ref{tab:result}, with their respective analysis results also presented in the table.
Model C is an intermediate work created by combining A and B, and Data D is the generated output from C. 
Work E involves republishing C as a service with the intent to sell, while Work F involves republishing D as a literary form, also with the intent to sell.
To be more convincing, we use real-world models and their respective licenses for demonstration.

\begin{table}[tbp]
    \caption{Settings and Corresponding Analysis Results from MG Analyzer with Fuzz Form Matching Enabled.}
    \scriptsize
    \begin{tabular}{|ll|}
    \hline \rowcolor[gray]{.8}
    \multicolumn{1}{|l|}{\textbf{Work: License}} & \textbf{Work: Report code} \\ \hline
    \multicolumn{1}{|p{3.2cm}|}{(i) C: AGPL-3.0. D: Unlicense. \newline (ii) D: Unlicense } & \multicolumn{1}{p{4.5cm}|}{ (i) E: N1N2N3$\times$2; \textbf{W5}$\times$2; W1$\times$3. F: W1$\times$3. \newline (ii) E: W1$\times$2. F: W1.} \\ \hline
    \multicolumn{2}{|p{8cm}|}{\textbf{OSS License Setting:} \newline
    (i) A$\leftarrow$PhoBERT~\cite{nguyen2020phobert}: AGPL-3.0. B$\leftarrow$CKIP-Transformers~\cite{lin2022hantrans}: GPL-3.0. \newline
    (ii) A$\leftarrow$PhoBERT~\cite{nguyen2020phobert}: AGPL-3.0. B$\leftarrow$\emph{None}.
    }\\ \hline \hline

    \multicolumn{1}{|p{3.2cm}|}{C: Unlicense. D: Unlicense.} & E: W1$\times$2; \textbf{E2}$\times$2; \textbf{E5}$\times$2. F: W1$\times$2 . \\ \hline
    \multicolumn{2}{|p{8cm}|}{\textbf{Free-content License Setting:}\newline
        A$\leftarrow$ MPT-Chat~\cite{MosaicML2023Introducing}: CC-BY-NC-SA-4.0.  B$\leftarrow$Command R+~\cite{cohereforai2024c4ai}: CC-BY-NC-4.0.
    } \\ \hline \hline

    \multicolumn{1}{|p{3.2cm}|}{(iii) D: Unlicense. \newline (iv) D: Unlicense.} & \multicolumn{1}{p{4.5cm}|}{ (iii) E: N1; N2; \textbf{W5} . F: \emph{None}. \newline (iv) E: N1; N2; W2$\times$2; \textbf{E2}; \textbf{E5}. F: W2, \textbf{E5}.} \\ \hline
    \multicolumn{2}{|p{8cm}|}{\textbf{Model License Setting:}\newline
        (iii) A$\leftarrow$ MG-BY-OS. B$\leftarrow$\emph{None}. (iv) A$\leftarrow$ MG-BY-NC. B$\leftarrow$\emph{None}.
    } \\ \hline
    \end{tabular}
    \label{tab:result}
\end{table}

First, we evaluate two OSS licenses: AGPL-3.0 and GPL-3.0, which are considered enforceable open-source licenses with copyleft clauses.
In setting (i), Work E triggers two \emph{Disclose Source Code} warnings (code W5, refer to Table~\ref{tab:report} for code definitions), and AGPL-3.0 successfully proliferates\footnote{Although two copyleft conditions are triggered simultaneously, they can be resolved since GPL-3.0 is compatible with AGPL-3.0. Appendix~\ref{apdx:grapher} visualizes this workflow.} to Work C.
However, with a small adjustment, we can circumvent these clauses by republishing the generated content rather than providing the model as a service. 
As demonstrated by Work D, there is no W5 warning, and the content is licensed under the Unlicense.
Furthermore, the condition in AGPL-3.0 that triggers the disclose code clauses related to remote access is \emph{you modify the Program}, which means you can directly republish copies as a service to circumvent this clause.
As reflected in setting (ii), there are no more W5 warnings, only two warnings related to the non-standard licensing remain.

Second, we evaluate two free-content licenses: CC-BY-NC-SA-4.0 and CC-BY-NC-4.0, both of which prohibit the commercial use of the governed work. 
As shown in the results, the republication of Work E successfully triggers E2 errors because the rights to commercial use are reserved. 
However, we can still circumvent these clauses by generating and then sharing the output, as these licenses lack rules regarding the generated work.

Third, we evaluate MG Licenses: MG-BY-OS and MG-BY-NC, which contain open sourcing and non-commercial use clauses, respectively.
In setting (iii), our MG-BY-OS license successfully triggers the W5 warning, indicating that Work E must disclose its source code.
In setting (iv), the non-commercial use error E5 is reported by the generated Work F.
As a model license, we do not enforce licensing on generated content, allowing Work D to be licensed under the Unlicense, albeit with certain restrictions.
A summary of the rights granted and restrictions imposed by these licenses can be found in their \emph{Model Sheet}, as previously mentioned.

It is worth mentioning that all results were obtained with \emph{fuzzy form matching} enabled, maximizing the detection of potential risks.
If these \emph{fuzz rules} were disabled, fewer issues would be reported. 
Furthermore, our MG Analyzer is designed to help developers be aware of potential compliance issues in ML projects. 
Its results should not be considered legal advice or a defense in dispute resolution. Please refer to our disclaimers in Appendix~\ref{apdx:disclamers}.
The most promising application lies in providing license analysis within complex ML workflows. An in-the-wild case can be found in Appendix~\ref{apdx:case2}.

\begin{tcolorbox}[boxrule=0.2mm, boxsep=1mm, left=0mm, right=0mm, top=0.1mm, bottom=0.1mm]
    \textbf{Summary}: GPL, AGPL, and CC licenses can be easily circumvented, leading to unintended misuse of the ML models they govern. In contrast, MG Licenses offer greater clarity and flexibility tailored to various publishing scenarios, promoting a more standardized and transparent approach to model licensing.
\end{tcolorbox}

\section{Conclusion}
\label{sec:conclusion}
Non-standard licensing is prevalent in ML projects, and the underlying risks are often neglected. 
To reveal these risks, we propose formal ontologies for describing ML workflows and develop the MG Analyzer to detect compliance issues based on it. 
To promote more standardized licensing in the future, we have drafted MG Licenses to provide flexible licensing solutions for model publishing. 
Our experiments show that commonly used OSS and CC licenses are unsuitable for model publishing, while MG Licenses provide a viable alternative.

\begin{acks}
This research is sponsored by Shanghai Pujiang Programme (Award No. 25PJA029).
This research is also supported in part by the National Research Foundation, Singapore and Infocomm Media Development Authority under its Trust Tech Funding Initiative. Any opinions, findings and conclusions or recommendations expressed in this material are those of the author(s) and do not reflect the views of National Research Foundation, Singapore and Infocomm Media Development Authority.
\end{acks}

\bibliographystyle{ACM-Reference-Format}
\balance
\bibliography{REF}

\appendix
\balance
\section{Limitations}
\label{apdx:disclamers}
The main limitation of MG Analyzer is that it is relied on accurate description of ML workflow production.
However, there is currently no standardized framework for capturing and organizing such information, which is typically scattered across Model Cards and Datasheets~\cite{biderman2022datasheet} in unstructured natural language form.
Furthermore, even model-sharing platforms like Hugging Face provide user-annotated metadata such as "trained on" and "merged from," but this information is often low quality and incomplete.
This limitation hinders the application of our analyzer to large-scale compliance analysis of real-world ML workflows and highlights the need for future efforts to build a comprehensive database that can trace the evolution of ML assets.
Another challenge arises from ambiguity in legal interpretations and dispute resolution. 
Take SmolLM-360M-Instruct in Figure~\ref{fig:case2} as an example. It is finetuned on data generated by Llama 3.1, which can be interpreted as a form of distillation, potentially rendering it a derivative work of Llama 3.1.
However, an opposing perspective argues that since Llama 3.1 was not directly used during training, the resulting model may not be subject to the terms of its license agreement.
This legal uncertainty limits the analyzer's ability to provide definitive compliance assessments in complex or indirect reuse scenarios.

There are also some critical voices regarding the Model Sheet included in our proposed MG Licenses, as noted by several legal experts.
While meant to aid understanding, including it in the main license text may create ambiguity and could encourage unauthorized modifications.
Currently, MG Licenses are being actively updated based on feedback, with several variants submitted to the OSI for approval.
Please see the disclaimers that follow.

\emph{\textbf{Disclaimers: The information in this article is for general informational purposes only and does not constitute legal advice. Views, opinions, and recommendations expressed are solely those of the author(s) and do not represent any organization. Do not rely on this material as a substitute for professional legal advice tailored to your specific circumstances.}}

\section{Encoded Rules and Reasoning Logic}
\label{apdx:rules}
In this section, we present the one encoded "derivative" rule for the GPL-3.0 license in Turtle format, followed by a snippet of the reasoning logic employed for license determination and rights granting analysis in Notation3.

\begin{tcolorbox}[title=MGLicenseRule.ttl > GPL-3.0-derivative-rule-1, colback=white, colframe=gray, boxrule=0.2mm, boxsep=1mm, left=0mm, right=0mm, top=0.1mm, bottom=0.1mm] 
  \begin{lstlisting}[language=Turtle, breaklines=false, columns=fullflexible, basicstyle=\footnotesize]
@prefix mg: <http://~/rdf/terms#> .
mg:GPL-3.0-derivative-rule-1 a mg:Rule ;
    mg:hasOutputDef mg:derivative ;
    mg:targetActionType mg:Combine ;
    mg:targetInputWorkForm mg:code ;
    mg:targetOutputWorkForm mg:code ;
    mg:relicense mg:compatible-license ;
    mg:hasPublishRestriction mg:include_license_restriction, 
        mg:include_notice_restriction, mg:disclose_self_restriction, 
        mg:state_changes_restriction, mg:gnu_freedom_restriction ;
    mg:allowSharing true .
  \end{lstlisting}
\end{tcolorbox}

\begin{tcolorbox}[title=rules\_ruling.n3 > License Determination Logic (snippet), colback=white, colframe=gray, boxrule=0.2mm, boxsep=1mm, left=0mm, right=0mm, top=0.1mm, bottom=0.1mm] 
  \begin{lstlisting}[language=Turtle, breaklines=false, columns=fullflexible, basicstyle=\footnotesize]
@prefix log: <http://www.w3.org/2000/10/swap/log#> .
@prefix list: <http://www.w3.org/2000/10/swap/list#> .
@prefix mg: <http://~/rdf/terms#> .
{ # CASE-3: There have been multiple rulings requiring the output work's license 
 # to be compatible, and these rulings have the same (compatible) license.
    ?outw a mg:Work .
    _:x log:notIncludes { ?outw mg:hasLicense ?li } .
    # If all relied works have a license, we can determine the license of this work
    ({?outw mg:hasReliedwork ?relw } { ?relw mg:hasLicense ?li}) log:forAllIn _:t . 
    # There is no ruling that does NOT allow relicensing (to exclude CASE-2)
    ( {?outw!mg:hasRuling!mg:hasRule mg:relicense ?reli } 
      {?reli log:notEqualTo mg:none-license} ) log:forAllIn _:t .
    # Collect all compatible relicensable licenses into a list
    ( ?li
        { 
            ?outw mg:hasRuling ?rling .
            ?rling!mg:hasRule mg:relicense ?rule .
            ?rule log:equalTo mg:compatible-license . # Compatible
            ?rling mg:hasLicense ?li . 
            # This ruling should bind a license which provides a compatible list
            ?li mg:hasCompatibleLicense ?clist .
        }
    ?li_list ) log:collectAllIn _:t . # (CASE-1 will yield a empty list here).
    ?li_list list:first ?li_1st .
    ?li_1st mg:hasCompatibleLicense ?compat_list_1st .
    # Check whether there exists a compatible license from the first list 
    # that also appears in all other (including itself) compatible lists.
    # Return true if there is a license included in all compatible list.
    _:x log:includes 
        {   
            ?cli list:in ?compat_list_1st .
            ( {?li_list!list:member mg:hasCompatibleLicense ?compat_list_other} 
             {?cli list:in ?compat_list_other} ) log:forAllIn _:t .
        } .
} => { ?outw mg:hasLicense ?cli } .
  \end{lstlisting}
\end{tcolorbox}

\begin{tcolorbox}[title=rules\_analysis\_granting.n3 > Right Reserved Error, colback=white, colframe=gray, boxrule=0.2mm, boxsep=1mm, left=0mm, right=0mm, top=0.1mm, bottom=0.1mm] 
  \begin{lstlisting}[language=Turtle, breaklines=false, columns=fullflexible, basicstyle=\footnotesize]
{ # Error [Right Reserved]. License reserves the right for your action.
    ?req a mg:Request .
    ?req!mg:grant mg:hasUsage ?req_usage . # There may be multiple mg:Usage.
    ?req mg:targetAction ?a .
    ?req mg:targetWork ?outw .
    ?req <- mg:hasRequest ?inw .
    ?inw mg:hasLicense ?li . # The target work must have a license.
    # Collect all reserved rights according to the license.
    ( ?r { ?li!mg:reserve mg:hasUsage ?r . } ?reserved_list ) log:collectAllIn _:x .
    ?req_usage list:in ?reserved_list .
    ( ?a ?inw ?outw ?req_usage ) log:skolem ?geniri . # Keep unique
    ("** Error ** [Right Reserved] " ?req_usage " is reserved by " ?li " license for " 
        ?a " action on " ?inw " to produce " ?outw) string:concatenation ?content .
} => {
    ?geniri a mg:Error ;
        mg:reportBy ?a ;
        mg:reportType "Right Reserved" ;
        mg:content ?content .
} .
  \end{lstlisting}
\end{tcolorbox}

\begin{table}[h]
    \caption{Supported Actions in MG Analyzer Following Rule Alignment. The symbols $=$ and $\approx$ indicate that the Type/Form of output work and input work are the same and may differ, respectively. The corresponding \textcolor{Construct}{OSS}, \textcolor{Ruling}{Free-content}, and \textcolor{Request}{Model} license terms for each action are listed. }
    \tiny
    \label{tab:action}
    \begin{tabular}{|l|c|c|p{3.4cm}|p{1.5cm}|}
    \hline
    \rowcolor[gray]{.8}
    Action & Type & Form & Composition & Terms \\ \hline

    Copy & $=$ & $=$ & Output and input are exactly \textbf{same}. & Copy Duplicate \\ \hline

    Combine & $\approx$ & $\approx$ & \textbf{Entire} input included in output. & \textcolor{Construct}{Link} \textcolor{Construct}{Aggregate} \textcolor{Request}{MoE} \textcolor{Ruling}{Arrange} \textcolor{Ruling}{Collect} \\ \hline

    Modify & $=$ & $=$ & Output includes a \textbf{significant} portion of input and \textbf{can be reverted}. & Modify \textcolor{Request}{Fine-tune}   \\ \hline

    Amalgamate & $=$ & $=$ & Output includes portions of input but \textbf{cannot be reverted}. & Modify \textcolor{Ruling}{Remix} \textcolor{Request}{Fusion} \\ \hline

    Train & $=$ & $=$ & Output has the \textbf{same structure} as input and may contain a \textbf{negligible} portion of it. & \textcolor{Ruling}{Alter Adapt} \textcolor{Request}{Train} \\ \hline

    Generate & $\approx$ & $\approx$ & Output does not contain any portion of input and may \textbf{perform differently} from it. & Output \textcolor{Request}{Generate} \newline \textcolor{Request}{Synthetic} \\ \hline

    Distill & $=$ & $=$ & Output does not contain any portion of input but \textbf{performs similarly} to it. & \textcolor{Request}{Distill Transfer} \newline \textcolor{Ruling}{Extract} \\ \hline

    Embed & $=$ & $=$ & The output contains no portion of the input, but a \textbf{mapping} converts input to output. & \textcolor{Ruling}{Translate Transform} \\ \hline

    Publish & $=$ & $\approx$ & Output is the \textbf{same} as the input but may have a \textbf{different form}.  & \textcolor{Construct}{Redistribute} \textcolor{Ruling}{Perform Display Disseminate} \\ \hline
    \end{tabular}
\end{table} 

\section{Supplementary of MG Licenses}
\label{apdx:model_sheet}

\begin{figure}[b]
  \centering
  \includegraphics[width=\linewidth]{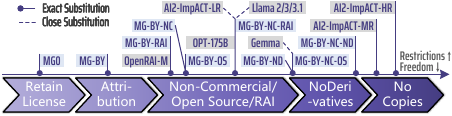}
  \caption{The Coverage of Model Publishing Scenarios by MG Licenses.}
  \Description{}
  \label{fig:license}
\end{figure}

Figure~\ref{fig:license} shows the coverage of MG Licenses. MG Licenses can serve as a substitute for all other model licenses in general model publishing purposes.
In particular, the MG-BY-RAI license can substitute for OpenRAIL-M; the MG-BY-NC-RAI license can be seen as a substitute for Llama 2/3/3.1, OPT-175B, and AI2-ImpaACT-LR licenses; and the MG-BY-OS license can nearly substitute for the Gemma license in general model publishing scenarios.

\section{Supplementary License Analysis Results: Complex In-the-Wild Workflow}
\label{apdx:case2}

\begin{figure}[h]
    \centering
    \includegraphics[width=\linewidth]{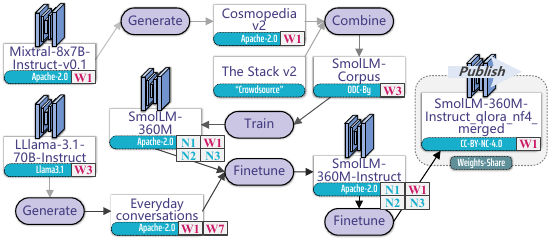}
    \caption{Analysis Results for an In-the-Wild Workflow based on SmolLM. Detected Issues Are Indicated in the Lower-Right Corner of Each Reported Work.}
    \label{fig:case2}
\end{figure}

\begin{figure*}[t]
  \centering
  \includegraphics[width=\linewidth]{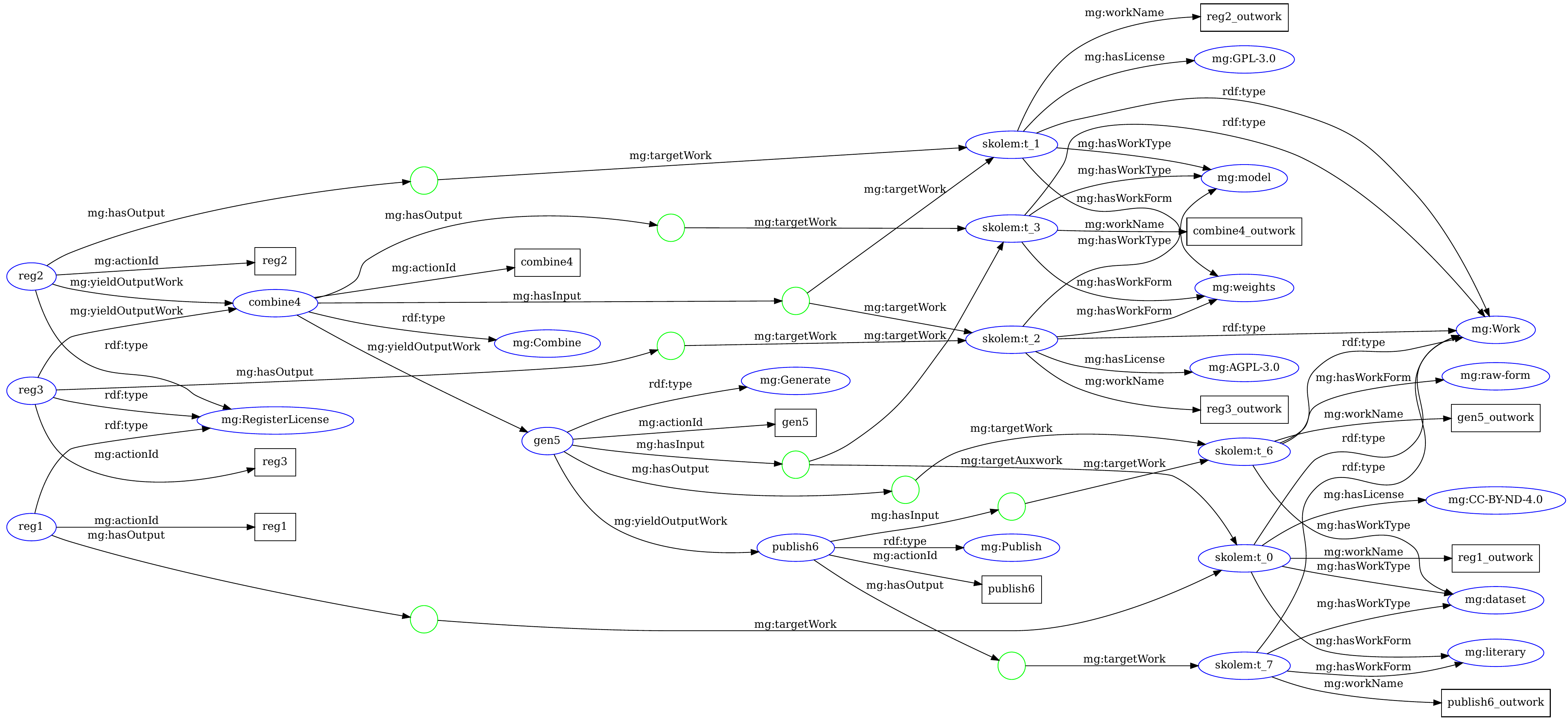}
  \caption{RDF Graph Visualization of the OSS Licenses Example Workflow in Section~\ref{sec:case}, Setting (i).}
  \Description{}
  \label{fig:grapher}
\end{figure*}

\begin{figure*}[t]
  \centering
  \includegraphics[width=\linewidth]{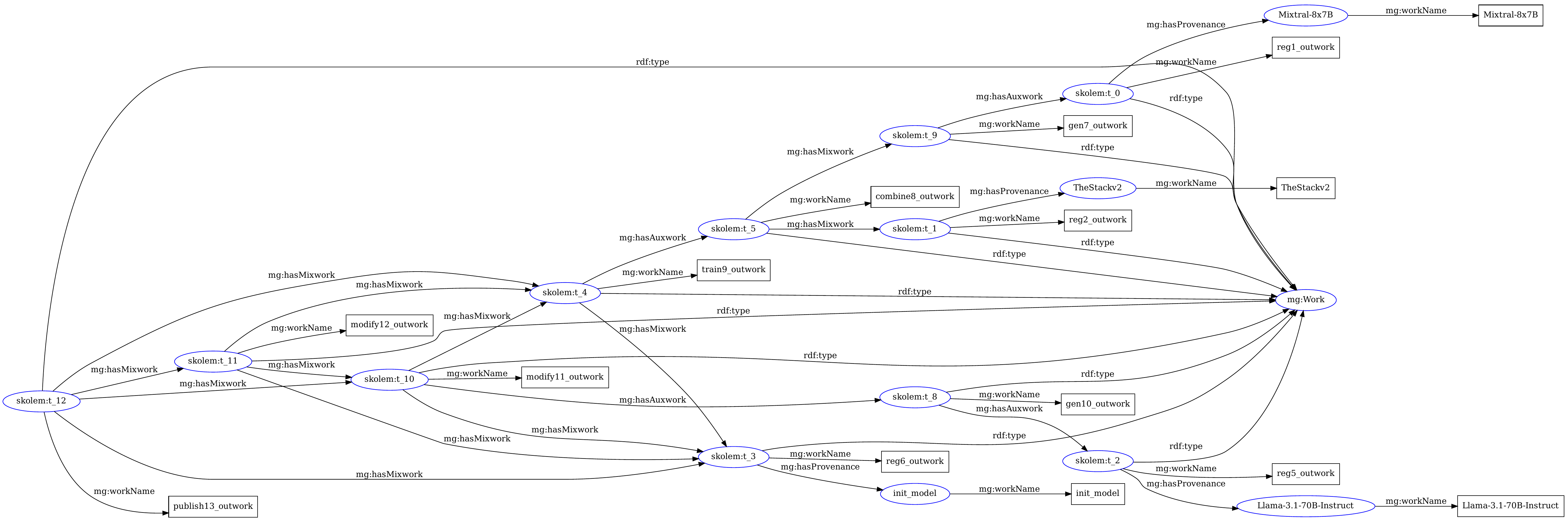}
  \caption{RDF Graph Visualization of the Compositional Dependency Reasoned by MG Analyzer for the SmolLM Example Workflow in Appendix~\ref{apdx:case2}.}
  \label{fig:grapher2}
\end{figure*}

While MG Analyzer is primarily designed for pre-deployment compliance analysis of ML workflows, it can also be applied to in-the-wild workflows, provided that licenses for intermediate assets are pre-assigned.
In this section, we present the analysis results of a SmolLM-based workflow~\cite{allal2024SmolLM}. 
Its reasoned compositional dependencies are shown in Figure~\ref{fig:grapher2}, and the production chain is illustrated in Figure~\ref{fig:case2}.
This complex workflow involves the generation of synthetic data for training and finetuning, as well as dataset combination.
Although most assets have been relicensed under permissive licenses, our analyzer still identifies six notices and six warnings. 
Note that, the Stack v2~\cite{lozhkov2024starcoder} is a crowdsourced dataset containing code under permissive or public domain licenses. In our analysis, we treat its license as Unlicense. For demonstration purposes, we assume this workflow is constructed and published in a single step, rather than through repeated construct-then-publish cycles.
The most commonly reported warning is \emph{License Type Mismatch}, which can lead to increased ambiguity in license interpretation and may raise the risk of misusing licensed works.
Another warning is triggered by Llama 3.1 and ODC-By, both categorized as \emph{Possibly Revocable Licenses}.
Warning W7 concerns the use of the \emph{everydata conversations} dataset, which must comply with the \emph{Use Behavior} restrictions specified in the Llama 3.1 license.
The six notices are triggered by finetuning on models licensed under Apache-2.0, which requires including the license file, providing proper notices, and stating any changes when redistributing derivative works.

It is worth emphasizing that legal noncompliance issues can become increasingly difficult to address as the reuse chain evolves. For example, \emph{Use Behavior} restrictions are often obscured through substream model reuse over multiple generations. The model cards~\cite{mitchell2019model} of \emph{Everyday Conversations} and \emph{SmolLM-360M-Instruct}, for instance, do not mention such constraints, posing a potential risk of unintended license violations.
Moreover, because detailed ML workflow metadata is not systematically recorded or standardized in current practice, it is challenging to obtain complete reproduction information for accurate analysis. This may compromise the comprehensiveness of our results.
Please refer to the Appendix~\ref{apdx:disclamers} for a discussion of our work's limitations.

\section{RDF Workflow Visualization}
\label{apdx:grapher}
Figure~\ref{fig:grapher} illustrates the workflow constructed by the MG Analyzer, as presented in the OSS licenses example (Section~\ref{sec:case}, Setting (i)) involving the publishing of Data D.
Figure~\ref{fig:grapher2} provides a visualization for the case discussed in Appendix~\ref{apdx:case2}.
Both figures were generated using RDF Grapher, a tool provided by the Linked Data Finland platform~\cite{hyvonen2014linked}. Note that, for clarity, only a subset of the properties is displayed.

\end{document}